\documentclass[twocolumn,showpacs,showkeys,preprintnumbers,amsmath,amssymb,nofootinbib]{revtex4}

\usepackage{graphicx}
\usepackage{dcolumn}
\usepackage{bm}

\usepackage{ulem} 
\usepackage[usenames]{color}

\def\be{\begin{eqnarray}}
\def\ee{\end{eqnarray}}
\def\bc{\begin{center}}
\def\ec{\end{center}}
\def\trm{\textrm}
\def\dsp{\displaystyle}
\def\prt{\partial}

\def\LamStar{\Lambda ^{\ast}}
\def\LamN{\Lambda N}
\def\SigzN{\Sigma ^{0} N}
\def\SigpmN{\Sigma ^{\pm} N}
\def\Lamp{\Lambda p}
\def\Sigzp{\Sigma ^{0} p}
\def\Sigpn{\Sigma ^{+} n}
\def\Lamn{\Lambda n}
\def\Sigzn{\Sigma ^{0} n}
\def\Sigmp{\Sigma ^{-} p}
\def\KbarN{\bar{K} N}
\def\piS{\pi \Sigma}

\def\etL{\eta \Lambda}

\begin{document}

\preprint{KUNS-2204, YITP-09-24}

\title{$\bm{\Lambda (1405)}$-induced non-mesonic decay in kaonic nuclei}

\author{T. Sekihara \email{sekihara@ruby.scphys.kyoto-u.ac.jp}}
\author{D. Jido$^{1}$}
\author{Y. Kanada-En'yo$^{1}$}

\affiliation{Department of Physics, Kyoto University, Kyoto 606-8502, Japan}
\affiliation{$^{1}$Yukawa Institute for Theoretical Physics, 
Kyoto University, Kyoto 606-8502, Japan}



\date{\today}

\begin{abstract}
Non-mesonic decay of kaonic nuclei is investigated under a $\Lambda (1405)$ 
doorway picture where the $\bar{K}$ absorptions in nuclei take place through 
the $\Lambda (1405)$ resonance. Calculating $\Lambda (1405) N \to YN$ 
transitions with one-meson exchange, we find that the non-mesonic 
decay ratio $\Gamma _{\Lambda N} / \Gamma _{\Sigma ^{0} N}$ depends strongly on 
the ratio of the couplings $\Lambda (1405)$-$\bar{K} N$ and 
$\Lambda (1405)$-$\pi \Sigma$. 
Especially a larger $\Lambda (1405)$-$\bar{K} N$ coupling leads to 
enhancement of the decay to $\Lambda N$. 
Using the chiral unitary approach for description of the $\bar{K} N$ 
amplitudes, we obtain $\Gamma _{\Lambda N} / \Gamma _{\Sigma ^{0} N} \approx 1.2$ 
almost independently of the nucleon density, and find the total 
non-mesonic decay width calculated in 
uniform nuclear matter to be 22 MeV at the normal density. 
\end{abstract}

\maketitle

\section{Introduction}
\label{sec:Introduction}

The in-medium properties of hadrons are key clues to understand 
finite density QCD. Meson-nucleus bound systems are good 
experimental objects to observe the in-medium properties of hadrons.
Anti-kaon ($\bar{K}$) bound states in nuclear systems (kaonic 
nuclei) are theoretically expected to exist due to strongly attractive 
interactions between the $\bar{K}$ and 
nucleon~\cite{Kishimoto:1999yj,Friedman:1999rh,Akaishi:2002bg}.
In spite of experimental efforts to search for the $\bar{K}$-nuclear bound 
systems~\cite{Kishimoto:2003jr},
there are no clear evidences observed yet.

For the observation of the kaonic nuclei in experiments, 
strength of $\bar{K}$ absorptions is a key property. 
With smaller $\bar{K}$ absorption widths than level spacings of the 
bound states, missing mass spectroscopy may be suitable to observe 
the kaonic bound states in kaon transfer reactions, 
such as $(K^{-}, \, N)$ with nuclear targets~\cite{Kishimoto:1999yj}
(see also Ref.~\cite{Yamagata:2006sv}). For 
larger widths, coincident observations of particles emitted from
the decay of the kaonic nuclei may be essential to pin down the $\bar{K}$ 
bound states, and the decay patterns tell us the properties of the bound
states. 

The decay of the kaonic nuclei in strong interactions is categorized into 
two processes; one is the mesonic process such as $\KbarN \to \pi Y$, 
and the other is the non-mesonic process such as $\KbarN N \to Y N$, 
where $Y$ denotes hyperon ($\Lambda$ or $\Sigma$). The 
non-mesonic decays have advantage in experimental observations, 
since signals from kaonic nuclei are readily distinguished from 
backgrounds and no extra mesons do not have to be detected. 
Therefore, systematic studies of the non-mesonic 
decay of the kaonic nuclei are desirable. Especially the ratios of 
the decay widths are interesting, since they will be insensitive to 
details of the production mechanism. 

The absorption of the $\bar{K}$ in nuclei may take place dominantly 
through the $\Lambda (1405)$ ($\LamStar$) resonance, owning to the presence 
of the $\LamStar$ just below the $\KbarN$ threshold. 
Namely the $\LamStar$ can be a doorway of the $\bar{K}$ absorptions 
in nuclei. The $\LamStar$ doorway picture is more probable,
in case that the $\Lambda (1405)$ is a quasi-bound state of 
$\KbarN$~\cite{Dalitz:1967fp,Hyodo:2008xr}, 
which has large $\KbarN$ components as almost real particles.
Recently it has been reported based on the $\KbarN$ quasi-bound picture 
of the $\Lambda(1405)$ in 
Refs.~\cite{Akaishi:2002bg,KanadaEn'yo:2008wm,Dote:2008hw} 
that also in few-body hadronic systems 
the $\KbarN$ correlation is so strong as to cause the significant 
component of the $\LamStar$ resonance. It has been also pointed out that 
the $\LamStar$ as the $\KbarN$ quasi-bound 
is not so compact compared with a typical hadronic 
size~\cite{Sekihara:2008qk,Akaishi:2002bg}. 
Therefore the strong $\KbarN$ correlations are expected to be 
responsible for the $\LamStar$-induced decays of the $\bar{K}$ in nuclei. 

In this paper, we study the non-mesonic decay of the kaonic 
nuclei under the $\LamStar$-induced picture, in which 
the decay process has the strong $\KbarN$ correlation in its initial 
state. For this purpose, we calculate the $\LamStar N \to YN$ process 
($\LamStar p \to \Lamp, \, \Sigzp , \, 
\Sigpn$ and $\LamStar n \to \Lamn , \, \Sigzn , \, \Sigmp$)
in uniform nuclear matter with a one-meson exchange approach.
We will see that the decay ratio $\Gamma _{\LamN} 
/ \Gamma _{\SigzN}$ depends strongly on $s$-wave $\LamStar$ couplings to 
$\KbarN$ and $\pi \Sigma$. 
Fixing the $\LamStar$ couplings by the $\KbarN$ scattering 
amplitudes calculated in the chiral unitary approach, 
we evaluate the non-mesonic decay widths of $\LamStar$ in 
nuclear medium as functions of the nucleon density. 

The non-mesonic decay widths of the $\LamStar$ in nuclei are calculated 
by $\LamStar N \to YN$.  In Sec.~\ref{sec:Formulation}, our 
formulation of the $\LamStar N \to YN$ transition 
rate is explained. We show the ratio of the $\LamStar N$ transitions 
to $\Lambda N$ and $\Sigma N$ as a function of the $\LamStar$ couplings 
to $\KbarN$ and $\pi \Sigma$ in Sec.~\ref{sec:General}. The 
non-mesonic decay widths are evaluated in nuclear matter as
functions of the nucleon density  in Sec.~\ref{sec:Width}.
Section~\ref{sec:Conclusion} is devoted to a summary.

\section{Formulation of $\bm{\LamStar N \to Y N}$ process}
\label{sec:Formulation}

In this section,  we formulate the $\LamStar N \to Y N$ 
transition process in a meson-exchange approach. 
Here $N$ denotes $p$ or $n$, and $Y$ is $\Lambda$ or $\Sigma$.
The $\LamStar N \to Y N$ transition is a possible elementary process 
of the non-mesonic decay of the $\bar{K}$ in nuclear medium. 
Summing up the $\LamStar N \to Y N$ transition rate in terms 
of the initial nucleon states, we estimate the non-mesonic 
decay width of the $\bar{K}$ in nuclear medium as a function
of the nucleon density in Sec.~\ref{sec:Width}.

The transition rate is given by the transition probability divided 
by time $\mathcal{T}$ as
\begin{equation}
\gamma _{YN}   \equiv 
 \frac{1}{\mathcal{T}} \frac{1}{\mathcal{V}^{2}} 
 \frac{1}{4} \sum _{\trm{spin}} 
 \int d\Phi_{2}
 |S - 1|^{2} ,
 \label{eq:trarate}
\end{equation}
with the $S$-matrix $S$ for the transition process given 
by the transition amplitudes $T_{YN}$ as $S=1-i
 (2 \pi)^{4} \delta ^{4} (P_{\LamStar} + P_{\trm{in}} - P_{N} - P_{Y}) 
 T_{YN}$ and the energy-momenta, $P_{\LamStar}$, $P_{\trm{in}}$ for 
the initial $\LamStar$ and nucleon $N$, and  
$P_{Y}$, $P_{N}$ for the outgoing hyperon $Y$ and nucleon $N$, respectively. 
The subscript ``$Y N$'' of $\gamma _{YN}$ and $T_{YN}$ represents 
the particles in the final state. 
The wavefunctions of the baryons have been introduced 
as plain waves normalized as unity in a spatial volume $\mathcal{V}$.
We have taken spin summation for the final state and spin average 
for the initial state. The phase space $d\Phi_{2}$ of the final state
is given by
\begin{equation}
  d\Phi_{2} \equiv \frac{d^{3} p_{Y}}{(2 \pi)^{3}} \frac{2 M_{Y}}{2 E_{Y}} 
  \frac{d^{3} p_{N}}{(2 \pi)^{3}} \frac{2 M_{N}}{2 E_{N}} ,
\end{equation}
with the normalization of the baryonic state as 
$\langle \bm{p}_{i} | \bm{p}^{\prime}_{i} \rangle = (2\pi)^{3} 
2 E_{i} \delta^{3}(\bm{p}_{i} - \bm{p}^{\prime}_{i}) /(2M_{i})$ 
and $E_{i}=\sqrt{M^{2}_{i}+ \bm{p}^{2}_{i}}$.
The masses of the $\LamStar$, nucleon and hyperon
are denoted by $M_{\LamStar}$, $M_{N}$ and $M_{Y}$, respectively. 
The transition rate depends on the center-of-mass energy $E_{\rm c.m.}$,
equivalently the initial nucleon momentum. 

Performing the phase space integral in Eq.~(\ref{eq:trarate}) in the 
$\LamStar$ rest frame, we obtain
\begin{align}
\gamma _{YN} ( p_{\trm{in}} ) 
& = \frac{1}{\mathcal{V}} \frac{1}{8 \pi} 
\sum _{\trm{spin}} 
\int _{-1}^{1} d \cos \theta _{N} | T_{YN} |^{2} 
\nonumber \\ & \phantom{= ~}
\times \frac{p_{N} M_{Y} M_{N}}
{E_{Y} + E_{N} -  p_{\trm{in}} \cos \theta _{N} E_{N} / p_{N}} , 
\label{eq:non-M_0}
\end{align}
where 
$P_{Y}^{\mu} = ( E_{Y} , \; \bm{p}_{Y})$, 
$P_{N}^{\mu} = ( E_{N} , \; \bm{p}_{N} )$,
$P_{\trm{in}}^{\mu} = ( E_{\trm{in}} , \; \bm{p}_{\trm{in}} )$ 
and $\theta_{N}$ denotes the angle of $\bm{p}_{N}$ to $\bm{p}_{\trm{in}}$.
We have used $ (2 \pi)^{4} \delta ^{4}(0) = \mathcal{TV}$.
The factor $1/\mathcal{V}$ is responsible 
for the fact that only one $N$ exists in the initial state 
in the volume $\mathcal{V}$.
Later it will be interpreted as nuclear density
in nuclear matter calculation.
In the case that the $\LamStar$ and $N$ interact in the initial state, 
the factor $1/\mathcal{V}$ for the plain-wave should be 
replaced by the square of a relative wavefunction of the 
two-body $\LamStar N$ system.

\begin{figure}[!t]
  \bc
  \begin{tabular*}{8.6cm}{@{\extracolsep{\fill}}cccc}
    \includegraphics[scale=0.14]{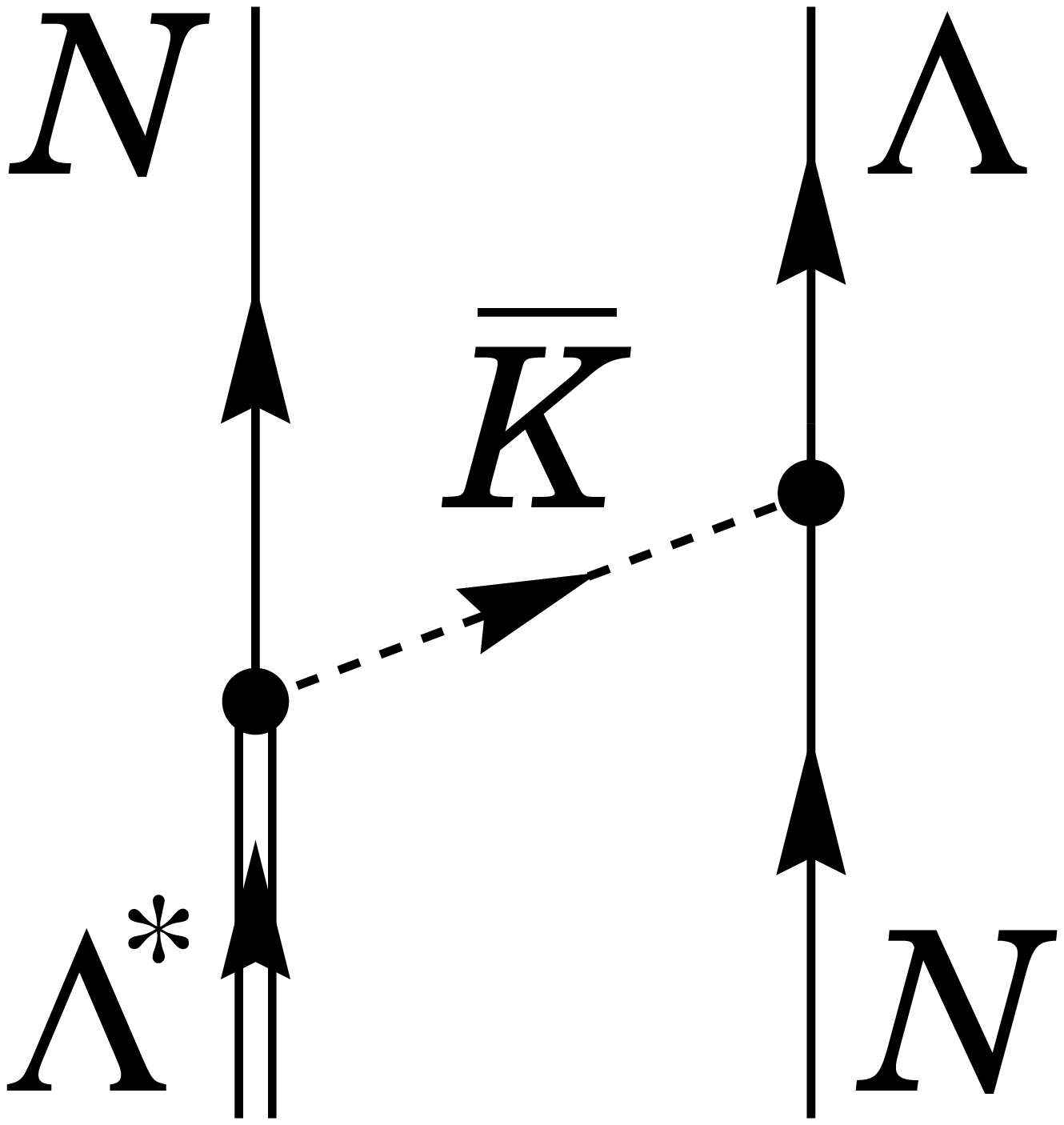}
    &
    \includegraphics[scale=0.14]{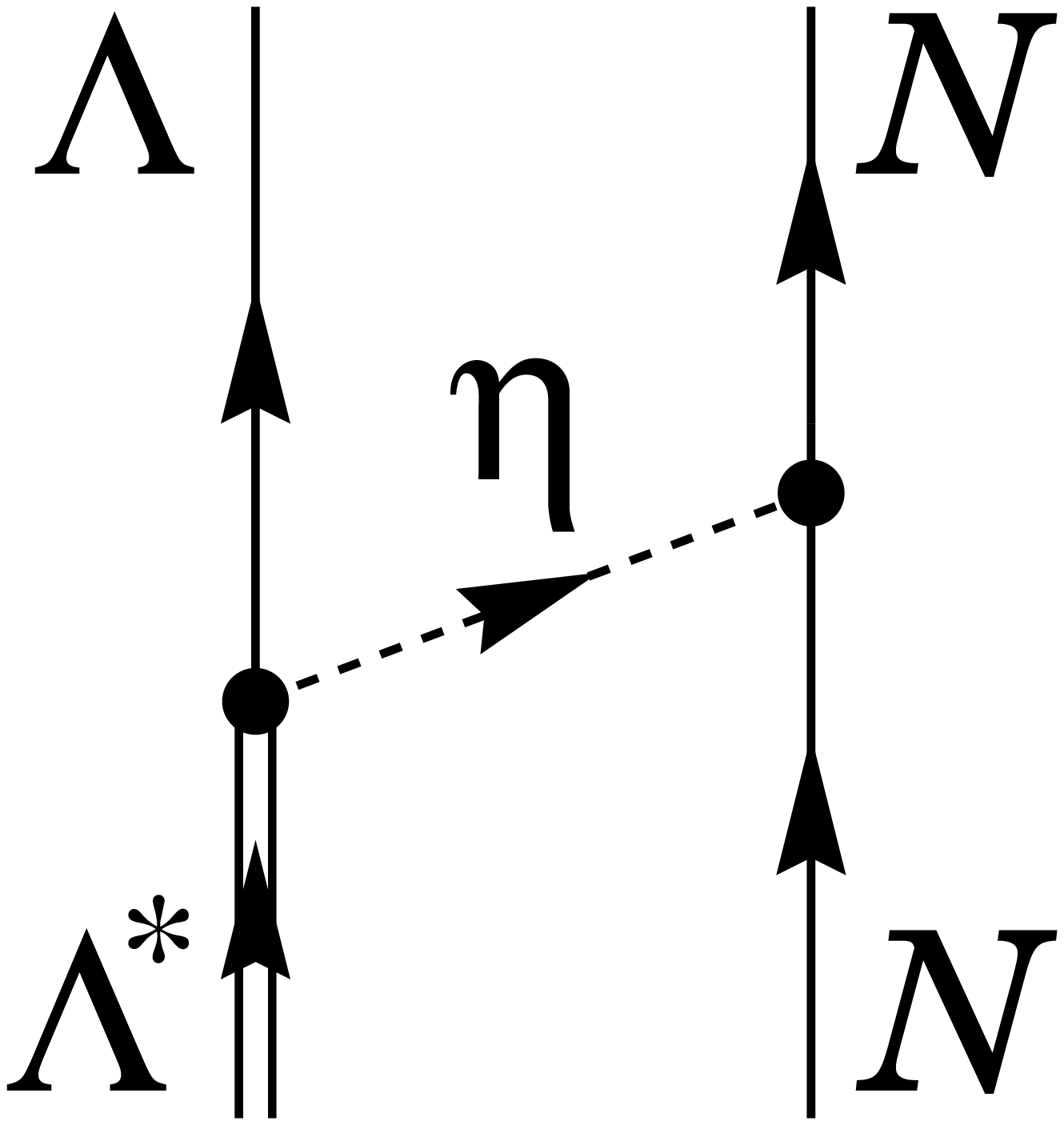} 
    &
    \includegraphics[scale=0.14]{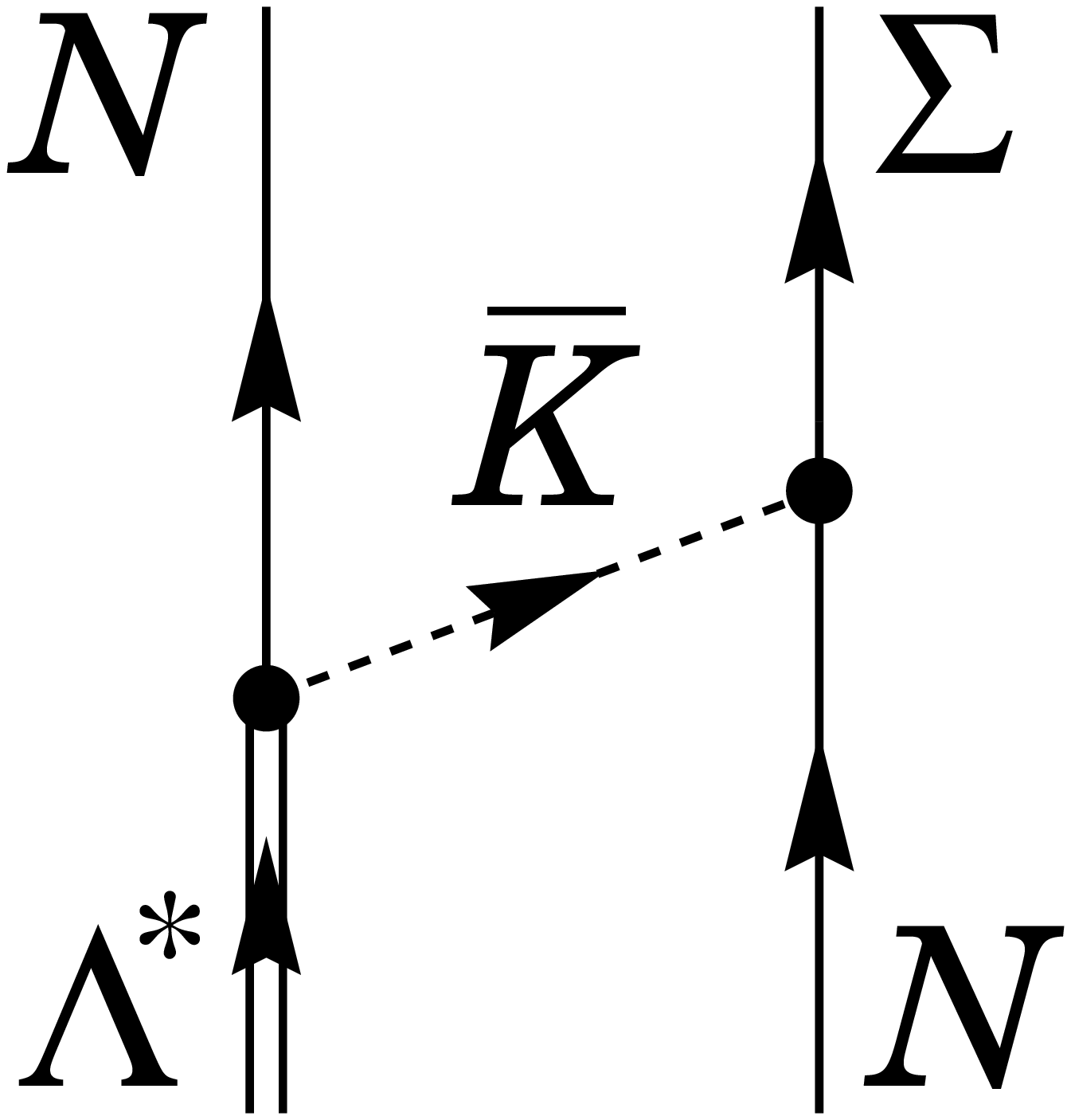}
    &
    \includegraphics[scale=0.14]{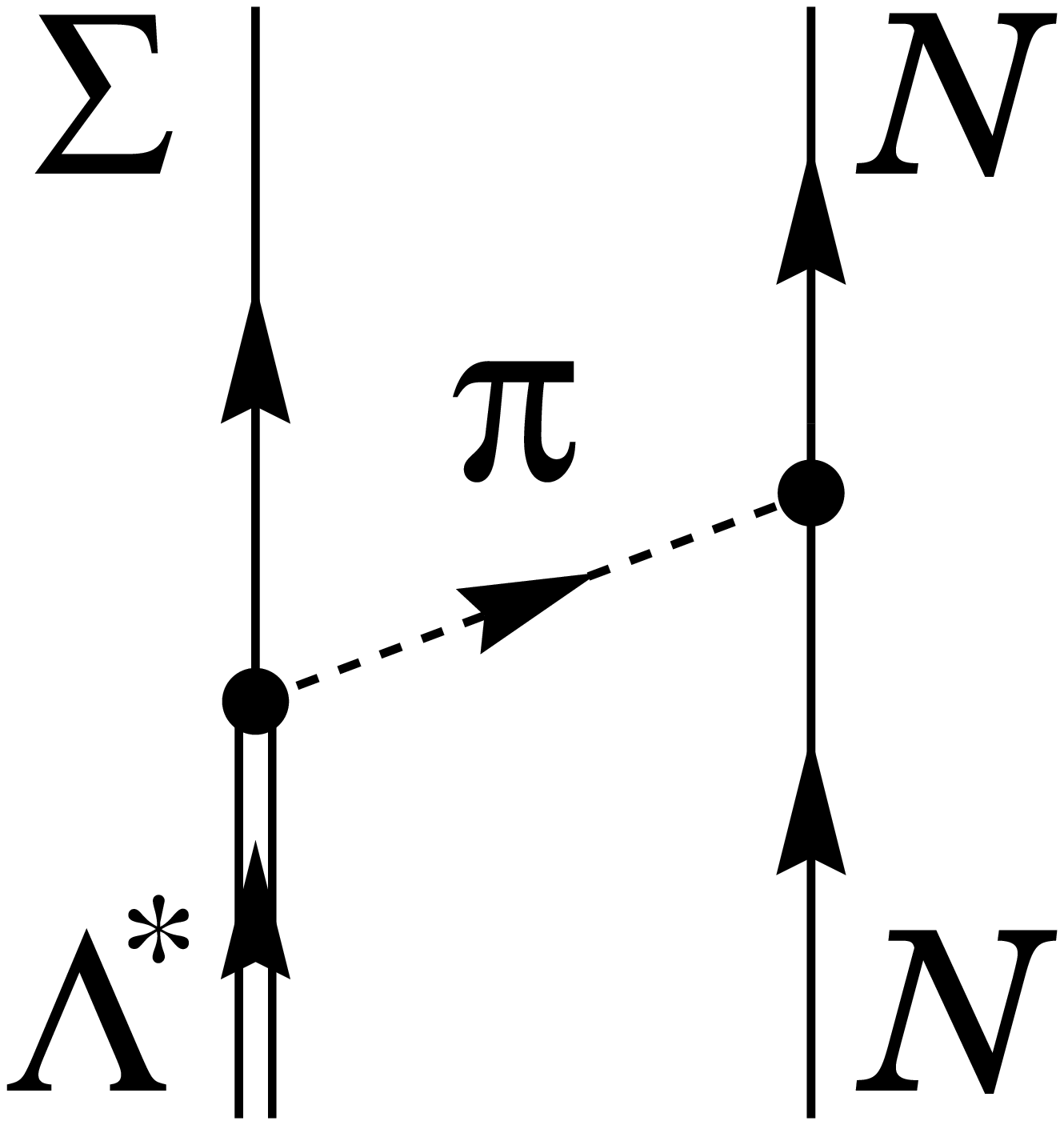} \\ 
    $\LamN 1$ & $\LamN 2$ &
    $\Sigma N 1$ & $\Sigma N 2$ 
  \end{tabular*}
  \ec
\vspace{-15pt} 
\caption{Feynman diagrams for the $\LamStar N \to YN$
  transition in the one-meson exchange model. 
  The upper two diagrams are for the $\Lambda N$ final state
  and the lower two are for the $\Sigma N$ state. }
\label{fig:non-M}
\vspace{-5pt} 
\end{figure}

We evaluate the transition amplitudes $T_{YN}$ 
with one-meson exchange diagrams shown in Fig.~\ref{fig:non-M}.
This approach of non-mesonic decay calculation 
with the one-meson exchange diagrams 
was applied for $\Lambda$ hypernuclei, 
reproducing the experimental values of the ratios $\Gamma _{n}/\Gamma _{p}$ 
for wide range of hypernuclei well~\cite{Jido:2001am}.
We assume isospin symmetry, thus the transition amplitudes 
$T_{YN}$ are independent from 
type of the nucleons in the initial state 
because the $\LamStar$ has isospin $I=0$. 

The transition amplitudes with the $\Lambda N$ and $\Sigma^{0} N$ in the 
final state can be written in terms of two parts
corresponding to the exchanged meson: 
\begin{align}
  & T_{\LamN} \equiv T_{\LamN 1} - T_{\LamN 2} , 
\label{eq:TLamN} \\
  & T_{\SigzN} \equiv T_{\SigzN 1} - T_{\SigzN 2} , 
\label{eq:TSigzN}
\end{align}
where the indices ($\LamN 1$ etc.) are given in Fig.~\ref{fig:non-M}. 
The relative sign between two amplitudes comes from
exchange of $Y$ and $N$ in the final state. 
The amplitude $T_{\SigpmN}$ can be obtained by the isospin 
relation: $T_{\SigpmN} = \sqrt{2} T_{\SigzN}$. 

Each amplitude of the diagrams given in Fig.~\ref{fig:non-M} is 
evaluated by introducing 
$s$-wave $\LamStar MB$ coupling constants $G_{MB}$, 
$MBB$ interactions $V_{MBB}$ and meson propagators $\Pi_{M}(q_{M}^{2})$ 
($M$ and $B$ represent the mesons and baryons in the particle-basis, 
respectively) as
\begin{align}
& T_{\LamN 1} = G_{\KbarN} 
\Pi _{\bar{K}}(q_{\bar{K}}^{2}) 
\, V_{\bar{K} N \Lambda} , 
\label{eq:TLamN1} \\
& T_{\LamN 2} = G_{\etL} 
\Pi _{\eta} (q_{\eta}^{2}) 
\, V_{\eta N N} ,  
\label{eq:TLamN2}
\end{align}
for the $\LamN$ final state, and 
\begin{align}
& T_{\SigzN 1} = G_{\KbarN} 
\Pi _{\bar{K}} (q_{\bar{K}}^{2}) 
\, V_{\bar{K} N \Sigma ^{0}} , 
\label{eq:TSigzN1} \\
& T_{\SigzN 2} = G_{\piS} 
\Pi _{\pi} (q_{\pi}^{2}) 
\, V_{\pi ^{0} N N} , 
\label{eq:TSigzN2}
\end{align}
for the $\SigzN$ final state. Because of the isospin symmetry, 
we have $G_{\KbarN}=G_{K^{-} p}=G_{\bar{K}^{0} n}$ and $G_{\pi \Sigma}
=G_{\pi ^{0} \Sigma ^{0}}=G_{\pi ^{\pm} \Sigma ^{\mp}}$. 

The $s$-wave $\LamStar MB$ coupling constants $G_{MB}$ are 
determined by the properties of the $\LamStar$. In the present work,
first we treat $G_{MB}$ as model parameters, and later
we fix them from the chiral unitary approach, which reproduces
the $\LamStar$ in meson-baryon dynamics. 

The meson-baryon-baryon three-point interactions $V_{MBB}$ 
are obtained from the low 
energy theorem in the flavor $\trm{SU(3)}$ symmetry. The interaction 
Lagrangian is written as 
\be
\mathcal{L}_{\rm int} = 
- \frac{D + F}{\sqrt{2}f} \left \langle \bar{B} \gamma ^{\mu} \gamma ^{5} 
\prt _{\mu} \Phi B \right \rangle 
- \frac{D - F}{\sqrt{2}f} \left \langle \bar{B} \gamma ^{\mu} \gamma ^{5} 
B \prt _{\mu} \Phi \right \rangle ,
\label{eq:LagMBB}
\ee
where $D$ and $F$ are the low energy constants which cannot be 
determined by the flavor symmetry, and $f$ is the meson decay constant. 
We use empirical values of $D+F=g_{A}=1.26$ and $D-F=0.33$, which 
reproduce the hyperon $\beta$ decays  observed in experiment,
and $f=f_{\pi}=93.0 \, \trm{MeV}$ commonly for all the mesons.
The matrices $B$ and $\Phi$ are the $\trm{SU(3)}$ expressions of the 
baryon and meson fields, respectively. From the Lagrangian~(\ref{eq:LagMBB}), 
we obtain the $p$-wave $MBB$ interactions in non-relativistic limit: 
\be
- i V_{MBB} = - 
\frac{d_{MBB} D + f_{MBB} F}{f} 
\bm{q}_{M} \cdot \bm{\sigma} ,
\ee
with the incoming meson momentum $\bm{q}_{M}$,
Pauli matrices $\bm{\sigma}$ for baryon spin
and the SU(3) Clebsch-Gordan coefficients $d_{MBB}$ and $f_{MBB}$. 

The propagator $\Pi _{M}(q_{M}^{2})$ for the meson $M$ is given by  
\be
\Pi _{M} (q_{M}^{2}) \equiv \frac{1}{q_{M}^{2} - m_{M}^{2}}.
\ee
The exchange meson momenta, $q_{\pi}^{\mu}$ for 
$\pi$, $q_{\eta}^{\mu}$ for $\eta$ and $q_{\bar{K}}^{\mu}$ for $\bar{K}$,
are fixed kinematically as 
\be
q_{\pi}^{\mu} = q_{\eta}^{\mu} = P_{\LamStar}^{\mu} - P_{Y}^{\mu},  \quad
q_{\bar{K}}^{\mu} = P_{\LamStar}^{\mu} - P_{N}^{\mu} .
\ee
In our study we take into account the short-range correlation, which 
compensates the short-range behavior of the meson-exchange 
potential~\cite{Oset:1979bi}. 
This can be introduced by changing the meson propagators to 
\be
\Pi (q^{2}) \to 
\tilde{\Pi} (q^{2}) \equiv 
\Pi (q^{2}) [F (q^{2})]^{2} - \Pi (\tilde{q}^{2}) [F (\tilde{q}^{2})]^{2} ,
\label{eq:Correlation}
\ee
with the mono-pole form factor 
$\dsp F(q^{2})= \frac{\Lambda ^{2}}{\Lambda ^{2} - q^{2}}$ 
and $\tilde{q}^{2} = q^{2} - q_{\trm{C}}^{2}$. We choose typical parameter 
set, 
$\Lambda = 1.0 \, \trm{GeV}$ and 
$q_{\trm{C}} = 780 \, \trm{MeV}$~\cite{Jido:2001am}.

\section{Ratio of transition rates
\protect\\ 
-- general properties}
\label{sec:General}

In the previous section, we have obtained the transition rates of 
$\LamStar N \to YN$
in Eq.~(\ref{eq:non-M_0}) together with the transition amplitudes
(\ref{eq:TLamN1})-(\ref{eq:TSigzN2}) in the one-meson exchange model. 
In the present calculation, 
the $MBB$ interactions $V_{MBB}$ are fixed by the flavor SU(3) symmetry 
with the empirical values of $D$ and $F$, while
the $\LamStar MB$ coupling constants $G_{MB}$ 
are to be determined by the properties of the $\LamStar$. 
In this section, we discuss $G_{MB}$ coupling dependence of
the ratio of the transition 
rates to $\Lambda N$ and $\Sigma^{0} N$, $\gamma _{\LamN} / 
\gamma _{\SigzN}$. 

First of all, it is helpful for understanding of the $\LamStar N \to YN$
transition rates to see which diagrams out of four shown in 
Fig.~\ref{fig:non-M} dominate the transition rates.
For this purpose, we show the $MBB$ coupling constants obtained 
by the low energy theorem with the present parameter $F$ and $D$:
\begin{align}
& 
V_{\bar{K} N \Lambda} \propto \frac{D + 3 F}{2 \sqrt{3}} \simeq 0.63, 
\quad 
V_{\eta N N} \propto \frac{D - 3 F}{2 \sqrt{3}} \simeq -0.10, 
\nonumber \\ 
& 
V_{\bar{K} N \Sigma ^{0}} \propto \frac{D - F}{2} \simeq 0.17, 
\quad 
V_{\pi ^{0} N N} \propto \frac{D + F}{2} \simeq 0.63. 
\nonumber 
\end{align}
These numbers indicate that the diagrams $\Lambda N1$ and $\Sigma N2$
give dominant contributions to the transition rates, while the 
diagram $\Lambda N2$, in which the $\eta$ meson is exchanged,
and the $\Sigma N1$ diagram are much less important with
factors 1/6 and 1/4 in the amplitudes than the dominant diagrams. 
Thus, the $\bar K$ exchange is important in the $\LamStar N \to
\Lambda N$ transition, whereas the $\pi$ exchange gives 
a dominant contribution in $\LamStar N \to \Sigma N$.

Let us calculate the ratio of the transition rates, 
$\gamma _{\LamN} / \gamma _{\SigzN}$. Here we assume 
$G_{\etL}=0$, since the $\eta$ exchange diagram is negligible 
due to the small $\eta NN$ coupling. Thus the ratio 
$\gamma _{\LamN} / \gamma _{\SigzN}$ becomes a function of  $G_{\KbarN}/G_{\piS}$.
We set also the initial nucleon momentum $ p_{\trm{in}} =0$ and 
the $\LamStar$ mass $M_{\LamStar}=1420 \, \trm{MeV}$\footnote{The chiral 
unitary model suggests that the resonance position 
of the $\Lambda (1405)$ in the $\bar KN$ channel is $1420 \, \trm{MeV}$ 
instead of $1405 \, \trm{MeV}$ 
and its width is 40 MeV, which is consistent with experimental 
data~\cite{Braun:1977wd}. See 
Ref.~\cite{Jido:2003cb,Jido:2009jf} for details.
}. 
Nevertheless, the transition rates are insensitive to 
$p_{\trm{in}}$ and $M_{\LamStar}$, because the phase space of the
$\LamStar N \to YN$ transition is sufficiently large against certain changes
of $p_{\trm{in}}$ and $M_{\LamStar}$, for instance, to 
$p_{\trm{in}} \approx 300 \, \trm{MeV}$ and 
$M_{\LamStar} \approx 1400 \, \trm{MeV}$.

\begin{figure}[!t]
  \bc
  \includegraphics[scale=0.25,angle=-90]{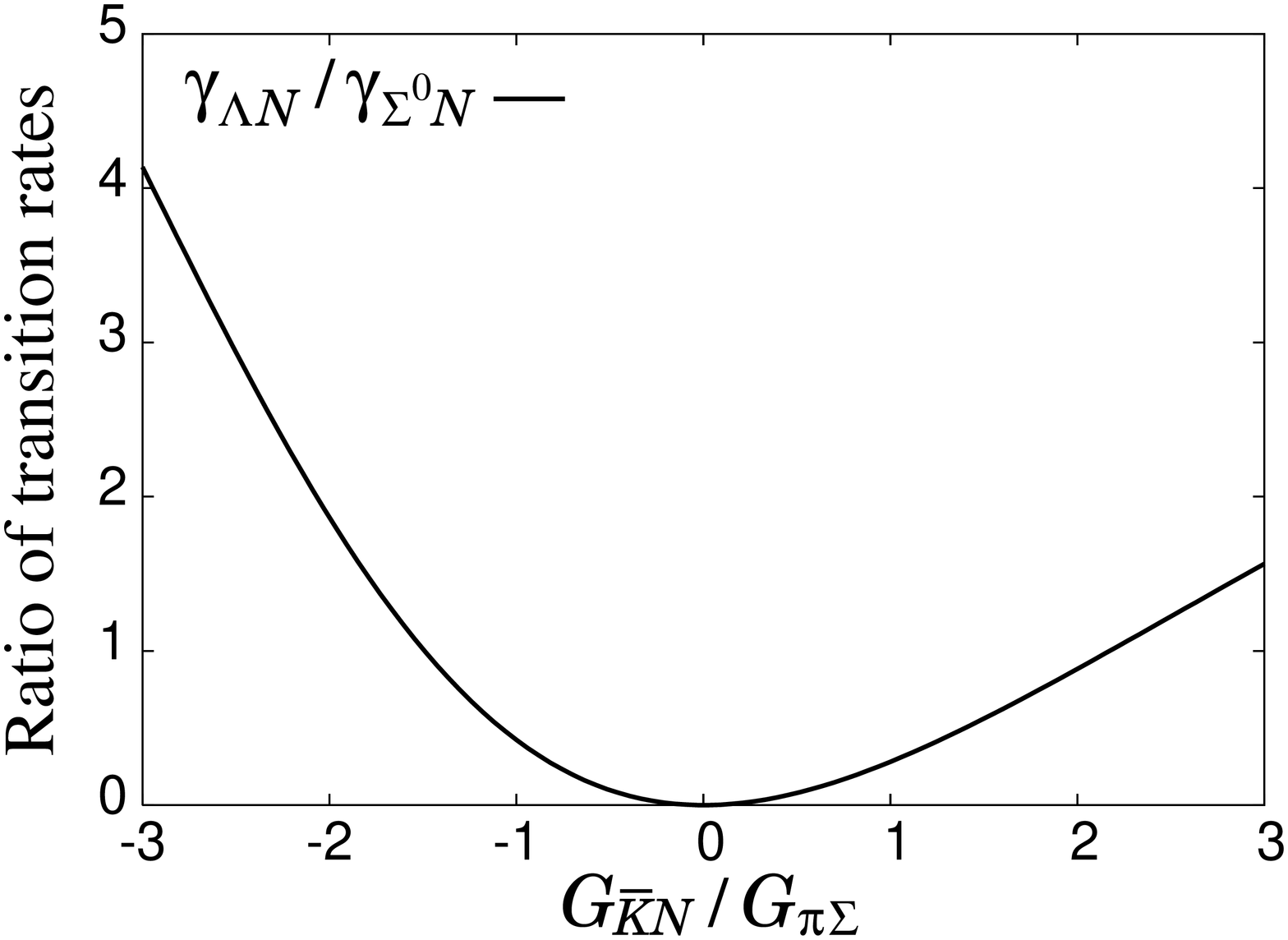}
  \ec
  \vspace{-10pt} 
\caption{Ratio of the transition rates of $\LamStar N$ to 
$\Lambda N$ and $\Sigma^{0} N$, $\gamma _{\LamN} / 
\gamma _{\SigzN}$, as a function of $G_{\KbarN}/G_{\piS}$.
We fix the initial nucleon momentum $ p_{\trm{in}} =0$ and 
the $\LamStar$ mass $M_{\LamStar}=1420 \, \trm{MeV}$
and assume  $G_{\etL}=0$.}
\label{fig:GratiovsCratio}
\vspace{-10pt} 
\end{figure}

In Fig.~\ref{fig:GratiovsCratio}, we show the numerical result of 
the ratio of the transition rates, $\gamma _{\LamN} / \gamma _{\SigzN}$,
as a function of $G_{\KbarN}/G_{\piS}$, in which we assume 
$G_{\KbarN}/G_{\piS}$ to be a real number. 
As seen in the figure, 
the ratio $\gamma _{\LamN} / \gamma _{\SigzN}$  has 
strong dependence on the coupling ratio $G_{\KbarN}/G_{\piS}$. 
This is because the transition to $\Lambda N$ is governed by
the $\bar K$ exchange and the transition to $\Sigma N$ 
is dominated by the $\pi$ exchange.
This result suggests that larger $\LamStar \KbarN$ 
coupling leads to enhancement of the decay ratio to $\LamN$ 
in the kaonic nuclei, although the $\LamStar$ 
in vacuum cannot decay into final states including $\Lambda$.

The asymmetric behavior of the ratio of the transition rates 
under change of the sign of the coupling ratio comes from 
interference between the diagrams $\Sigma N1$ and $\Sigma N2$.
$\Sigma N1$ gives subdominant contributions to $\gamma_{\Sigma N}$
and depends on $G_{\bar KN}$. If the relative sign of $G_{\bar KN}$ and
$G_{\pi \Sigma}$ is negative, the diagrams $\Sigma N1$ and $\Sigma N2$
are summed deconstructively. Thus, the ratio 
$\gamma _{\LamN} / \gamma _{\SigzN}$ for 
$G_{\KbarN}/G_{\piS} < 0$ is larger than the corresponding 
ratio with $G_{\KbarN}/G_{\piS} > 0$. 

Finally it is worth noting again that the ratio of the transition rates
is insensitive to the mass shift of the $\LamStar$. This means that, 
even in the case of in-medium modification of the $\LamStar$ mass,
as long as the $\LamStar$ resonance is predominant in the $\bar KN$
channel, the enhancement of the decay to $\Lambda N$ in the kaonic 
nuclei stems from larger coupling of the $\LamStar$ to $\bar KN$.
It is also noted that, even if the coupling constants $G_{\KbarN}$ 
and $G_{\piS}$ are complex numbers, 
one can obtain a similar behavior for $|G_{\KbarN}/G_{\piS}|$,
because the ratio of the transition rates is dominated by the two
diagrams, $\Lambda N1$ and $\Sigma N2$, which cannot interfere.

\section{Non-mesonic decay width of $\bm{\LamStar}$\protect\\
 in nuclear medium}
\label{sec:Width}

We calculate the non-mesonic decay of the $\bar{K}$ in uniform 
nuclear matter induced by the $\LamStar N \to YN$ transition ($N=p$ or $n$),
under the free Fermi gas approximation for nuclear matter 
in which non-interacting nucleons fill the momentum states 
up to the Fermi momentum $k_{\trm F}$. Assuming that the $\LamStar$ is at rest 
in nuclear medium, we evaluate the non-mesonic decay width 
by summing up the transition rate of $\LamStar N \to YN$
for all the pair of the $\LamStar$ and the nucleons in a unit volume 
$\mathcal{V}$:
\begin{equation}
\Gamma _{YN}  \equiv \sum _{i=1}^{A_{N}} 
\gamma _{YN} (k_{i}) , 
\end{equation}
where $A_{N}$ ($N = p$ or $n$) is the numbers of the protons or 
neutrons in $\mathcal V$ given by the Fermi momentum 
$k_{\trm{F} N}$ as
\be
A_{N}  = \frac{k_{\trm{F} N}^{3}}{3 \pi ^{2}} 
\mathcal{V} .
\ee
The proton and neutron densities are given by 
$\rho_{N} \equiv k_{\trm{F} N}^{3} / (3 \pi^{2})$. 
With the explicit expression of $\gamma_{YN}$ given in 
Eq.~(\ref{eq:non-M_0}) and replacing the summation over the nucleon 
numbers to integral with respect to the nucleon momentum, we obtain
\begin{align}
\Gamma _{YN} &= 
\int _{0}^{k_{\trm{F} N}} d k \frac{k^{2}}{8 \pi ^{3}} 
\sum _{\trm{spin}}
\int _{-1}^{1} d \cos \theta _{N} |T _{YN}|^{2} 
\nonumber \\ & \phantom{= ~}
\times \frac{p_{N} M_{Y} M_{N}}
{E_{Y} + E_{N} - k \cos \theta _{N} E_{N} / p_{N}} . 
\label{eq:GammaYN}
\end{align}
Here we note that, in the two-nucleon absorption of the $\bar{K}$, 
the emitted nucleon is not affected by Pauli blocking in nuclear medium, 
since the outgoing nucleon momentum is about 
$500 \, \trm{MeV}$ in the $\LamStar$-rest frame, which is much larger than 
the Fermi momentum for the nuclear saturation density. 

In the previous section, we have regarded the $\LamStar$ counplings 
$G_{MB}$ as free parameters. Here we estimate 
the $\LamStar MB$ coupling constants by the $s$-wave
$\KbarN$ scattering amplitudes with $I=0$ around the
$\LamStar$ resonance energy, which are dominated
by the $\LamStar$ contribution initiated by $\KbarN$.
Since the scattering amplitudes have the $\LamStar$ propagator
and the initial $\KbarN \to \LamStar$ coupling, we take
the ratios of the amplitudes to cancel out the propagator and 
the coupling for the extraction of the $\LamStar MB$ couplings: 
\be
\frac{G_{\KbarN}}{G_{\piS}} 
= \frac{t_{\KbarN} (W)}{t_{\piS} (W)} ,  \quad 
\frac{G_{\KbarN}}{G_{\etL}} 
= \frac{t_{\KbarN} (W)}{t_{\etL} (W)} , 
\label{eq:GoverG}
\ee
where $t_{MB}(W)$ is the scattering amplitude of $\KbarN (I=0)$ to 
$MB (\trm{particle-basis})$ 
with the center-of-mass energy $W$ chosen to be the $\LamStar$
resonance position on the real energy axis. 

For the description of the $\KbarN$ scattering amplitudes $t_{MB}(W)$,
we use the chiral unitary approach (ChUA), which reproduces 
well the $s$-wave scattering amplitudes of $\KbarN \to M B$
in the coupled-channels method based on chiral dynamics and in which 
the $\LamStar$ resonance is dynamically generated in the meson-baryon
scattering.
The model parameters of ChUA are determined so as to reproduce
the threshold properties of $K^{-} p$ observed in $K^{-}$ absorptions 
of kaonic hydrogen~\cite{Tovee:1971ga,Nowak:1978au}. The details are 
given in Ref.~\cite{Sekihara:2008qk}, which is based on 
Refs.~\cite{Oset:2001cn,Hyodo:2002pk}. 
Since the $\LamStar$ resonance position in the $\KbarN$ channel 
is 1420 MeV in ChUA, we take $W=M_{\LamStar}=1420 \, \trm{MeV}$
in Eq.~\eqref{eq:GoverG}.
In the present work, we use the $\KbarN$ scattering amplitude obtained
in vacuum. Extensions with in-medium $\LamStar$ modification 
are straightforward, as long as 
the $\LamStar$ doorway picture of the $\bar{K}$ absorption 
is valid in nuclear media. In-medium $\LamStar$ coupling constants can be 
obtained through Eq.~\eqref{eq:GoverG} with in-medium $\bar KN$ scattering
amplitudes as calculated, for instance,  
in Ref.~\cite{Waas:1996xh}. 

\begin{figure}[!t]
  \bc
  \includegraphics[scale=0.25,angle=-90]{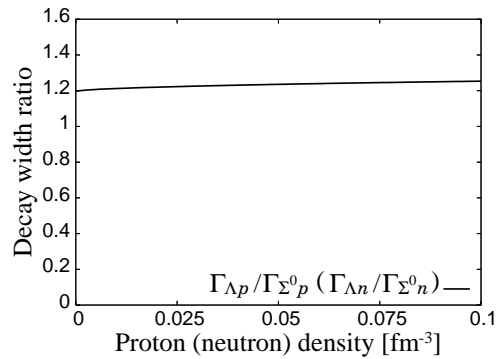}
  \ec
  \vspace{-10pt} 
\caption{Non-mesonic decay width ratio of kaonic nuclei.}
\label{fig:GratiovsRho}
\vspace{-10pt} 
\end{figure}

Fixing the ratios of the $\LamStar$ coupling constants, $G_{MB}$, 
we plot the ratio of the non-mesonic decay widths of the 
in-medium $\bar{K}$ 
to $\Lambda N$ and $\Sigma^{0}N$ as a function of the proton (or neutron) 
density in Fig.~\ref{fig:GratiovsRho}.
This figure shows that the ratio of the non-mesonic decay widths 
$\Gamma _{\LamN} / \Gamma _{\SigzN}$ is around $1.2$ 
almost independently of the nucleon density. 
The value $\Gamma _{\LamN} / \Gamma _{\SigzN} \approx 1.2$
is a consequence of the larger $\LamStar$ coupling to $\bar KN$
than $\pi \Sigma$. In ChUA, the ratio of the couplings 
$|G_{\KbarN} / G_{\piS}|$ is obtained as 2.5. 
The density independence of the ratio of the decay widths is 
due to the small dependence of the transition rate on the initial 
nucleon momentum, which is caused by sufficiently large phase space 
in the final states. It is worth noting that the ratios of the decay width
will be insensitive to details of the production mechanism of
the kaonic bound state, since they will be cancelled out. 

\begin{figure}[!t]
  \bc
  \includegraphics[scale=0.25,angle=-90]{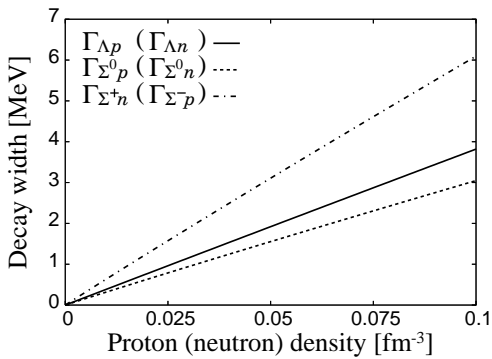}
  \ec
  \vspace{-10pt} 
\caption{Non-mesonic decay width of kaonic nuclei.}
\label{fig:GvsRho}
\vspace{-10pt} 
\end{figure}

To obtain absolute values of the non-mesonic decay width
in nuclear matter, we use the mesonic decay width
to $\pi \Sigma$ as a reference: 
\be
\Gamma _{\LamStar \to \piS} = 3 \times 
\frac{p_{\trm{CM}} M_{\Sigma}}{2 \pi M_{\LamStar}} |G_{\piS} |^{2} , 
\label{eq:normalization}
\ee
where $p_{\trm{CM}}$ is the center-of-mass momentum of $\pi$.
Taking the mesonic decay width being $\Gamma _{\LamStar \to \piS} =40\,
\trm{MeV}$ with $M_{\LamStar} =1420 \, \trm{MeV}$ as observed in a 
$K^{-}$ initiated channel in vacuum~\cite{Braun:1977wd,Jido:2009jf},
we obtain $|G_{\piS}| = 0.78$. The $G_{\piS}$ can be evaluated also
within ChUA as $|G_{\piS}|_{\trm{ChUA}}=0.78$ on the real axis 
using the Breit-Wigner formula, 
and $|G_{\piS}|_{\trm{pole}}=0.88$ at the $\LamStar$ complex pole
position~\cite{Jido:2003cb}. 

We show the non-mesonic decay widths calculated in nuclear 
medium, $\Gamma _{\LamN}$, $\Gamma _{\SigzN}$ and $\Gamma _{\SigpmN}$, 
in Fig.~\ref{fig:GvsRho}. The $\LamStar$ couplings are determined 
by Eq.~(\ref{eq:GoverG}) with the ChUA amplitudes, and are normalized by 
Eq.~(\ref{eq:normalization}) with $\Gamma _{\LamStar \to \piS} =40\, \trm{MeV}$.
The linear dependence of the decay widths is caused by insensitivity of the 
elementary transition rate $\gamma_{YN}$ to the nucleon momentum. 
At the normal nuclear density ($\rho _{\trm{B}} = 0.17 \, \trm{fm}^{-3}$),
the total non-mesonic decay width is $22 \, \trm{MeV}$, 
in which $\Gamma _{\Lamp} 
+ \Gamma _{\Lamn} = 7 \, \trm{MeV}$, $\Gamma _{\Sigzp} + \Gamma _{\Sigzn} 
= 5 \, \trm{MeV}$ and $\Gamma _{\Sigpn} + \Gamma _{\Sigmp} = 10 \, \trm{MeV}$. 
This is almost half of the mesonic decay width of $\LamStar$ 
($\sim 40 \, \trm{MeV}$). 

\vspace{40pt}

\section{Conclusion}
\label{sec:Conclusion}
We have investigated the non-mesonic decay of kaonic nuclei under the 
$\Lambda (1405)$ doorway picture in which the $\bar K$ absorption in 
nuclear medium takes place via the $\LamStar$ resonance 
through $\KbarN \to \LamStar$. 
Calculating the $\LamStar N \to YN$ transition rates in the 
one-meson exchange processes, we have found that the transition rate
to $\Lambda N$ is governed by the $\bar{K}$ exchange,
and consequently the ratio of the $\LamStar N$ transition rates 
to $\Lambda N$ and $\Sigma N$ strongly depends on the ratio
of the $\LamStar$ couplings, $G_{\bar KN}/G_{\pi\Sigma}$. 
Particularly, larger $\LamStar$ couplings to $\KbarN$
lead to enhancement of the non-mesonic two-body $\bar{K}$ absorption with 
$\Lambda N$ emission. 

Evaluating the $\LamStar$ couplings by the $\KbarN$ scattering 
amplitudes obtained by the chiral unitary approach, we have calculated
the widths of the non-mesonic decay of kaonic nuclei induced by 
$\LamStar N \to YN$. We have obtained the ratio of the non-mesonic decay 
widths to $\Lambda N$ and $\Sigma ^{0} N$ as 
$\Gamma _{\LamN} / \Gamma _{\SigzN} = 1.2$ almost independently 
of the nucleon density. We have also estimated that the total 
non-mesonic decay width is $22 \, \trm{MeV}$ at the saturation density. 
These findings will help us to understand the properties of the kaonic nuclei.

\begin{acknowledgements}
We would like to thank Dr.\ Fujioka for useful discussions. 
This work was supported in part by
the Grant-in-Aid for Scientific Research
from MEXT and JSPS (Nos.
   18540263,        
   20028004    	
   and 20540273), 
and the Grant-in-Aid for the Global COE Program 
"The Next Generation of Physics, Spun from Universality and Emergence" 
from MEXT of Japan.
This work was done under
the Yukawa International Program for Quark-hadron Sciences.

\end{acknowledgements}


\begin{thebibliography}{28}
\expandafter\ifx\csname natexlab\endcsname\relax\def\natexlab#1{#1}\fi
\expandafter\ifx\csname bibnamefont\endcsname\relax
  \def\bibnamefont#1{#1}\fi
\expandafter\ifx\csname bibfnamefont\endcsname\relax
  \def\bibfnamefont#1{#1}\fi
\expandafter\ifx\csname citenamefont\endcsname\relax
  \def\citenamefont#1{#1}\fi
\expandafter\ifx\csname url\endcsname\relax
  \def\url#1{\texttt{#1}}\fi
\expandafter\ifx\csname urlprefix\endcsname\relax\def\urlprefix{URL }\fi
\providecommand{\bibinfo}[2]{#2}
\providecommand{\eprint}[2][]{\url{#2}}

\bibitem[{\citenamefont{Kishimoto}(1999)}]{Kishimoto:1999yj}
\bibinfo{author}{\bibfnamefont{T.}~\bibnamefont{Kishimoto}},
  \bibinfo{journal}{Phys. Rev. Lett.} \textbf{\bibinfo{volume}{83}},
  \bibinfo{pages}{4701} (\bibinfo{year}{1999}).

\bibitem[{\citenamefont{Friedman and Gal}(1999)}]{Friedman:1999rh}
\bibinfo{author}{\bibfnamefont{E.}~\bibnamefont{Friedman}} \bibnamefont{and}
  \bibinfo{author}{\bibfnamefont{A.}~\bibnamefont{Gal}},
  \bibinfo{journal}{Phys. Lett.} \textbf{\bibinfo{volume}{B459}},
  \bibinfo{pages}{43} (\bibinfo{year}{1999}).

\bibitem[{\citenamefont{Akaishi and Yamazaki}(2002)}]{Akaishi:2002bg}
\bibinfo{author}{\bibfnamefont{Y.}~\bibnamefont{Akaishi}} \bibnamefont{and}
  \bibinfo{author}{\bibfnamefont{T.}~\bibnamefont{Yamazaki}},
  \bibinfo{journal}{Phys. Rev.} \textbf{\bibinfo{volume}{C65}},
  \bibinfo{pages}{044005} (\bibinfo{year}{2002});
%
\bibinfo{author}{\bibfnamefont{T.}~\bibnamefont{Yamazaki}} \bibnamefont{and}
  \bibinfo{author}{\bibfnamefont{Y.}~\bibnamefont{Akaishi}},
  \bibinfo{journal}{Proc. Japan Acad.} \textbf{\bibinfo{volume}{B83}},
  \bibinfo{pages}{144} (\bibinfo{year}{2007}).

\bibitem[{\citenamefont{Kishimoto et~al.}(2003)}]{Kishimoto:2003jr}
\bibinfo{author}{\bibfnamefont{T.}~\bibnamefont{Kishimoto}}
  \bibnamefont{et~al.}, \bibinfo{journal}{Prog. Theor. Phys. Suppl.}
  \textbf{\bibinfo{volume}{149}}, \bibinfo{pages}{264} (\bibinfo{year}{2003});
%
\bibinfo{author}{\bibfnamefont{M.}~\bibnamefont{Agnello}} \bibnamefont{et~al.}
  (\bibinfo{collaboration}{FINUDA}), \bibinfo{journal}{Phys. Rev. Lett.}
  \textbf{\bibinfo{volume}{94}}, \bibinfo{pages}{212303}
  (\bibinfo{year}{2005});
%
\bibinfo{author}{\bibfnamefont{T.}~\bibnamefont{Kishimoto}}
  \bibnamefont{et~al.}, \bibinfo{journal}{Nucl. Phys.}
  \textbf{\bibinfo{volume}{A754}}, \bibinfo{pages}{383} (\bibinfo{year}{2005});
%
\bibinfo{author}{\bibfnamefont{T.}~\bibnamefont{Suzuki}} \bibnamefont{et~al.},
  {\it ibid.}
  \textbf{\bibinfo{volume}{A754}},
  \bibinfo{pages}{375} (\bibinfo{year}{2005}).

\bibitem[{\citenamefont{Yamagata and Hirenzaki}(2007)}]{Yamagata:2006sv}
\bibinfo{author}{\bibfnamefont{J.}~\bibnamefont{Yamagata}} \bibnamefont{and}
  \bibinfo{author}{\bibfnamefont{S.}~\bibnamefont{Hirenzaki}},
  \bibinfo{journal}{Eur. Phys. J.} \textbf{\bibinfo{volume}{A31}},
  \bibinfo{pages}{255} (\bibinfo{year}{2007});
%
\bibinfo{author}{\bibfnamefont{J.}~\bibnamefont{Yamagata}},
  \bibinfo{author}{\bibfnamefont{H.}~\bibnamefont{Nagahiro}}, \bibnamefont{and}
  \bibinfo{author}{\bibfnamefont{S.}~\bibnamefont{Hirenzaki}},
  \bibinfo{journal}{Phys. Rev.} \textbf{\bibinfo{volume}{C74}},
  \bibinfo{pages}{014604} (\bibinfo{year}{2006}).

\bibitem[{\citenamefont{Dalitz et~al.}(1967)\citenamefont{Dalitz, Wong, and
  Rajasekaran}}]{Dalitz:1967fp}
\bibinfo{author}{\bibfnamefont{R.~H.} \bibnamefont{Dalitz}},
  \bibinfo{author}{\bibfnamefont{T.~C.} \bibnamefont{Wong}}, \bibnamefont{and}
  \bibinfo{author}{\bibfnamefont{G.}~\bibnamefont{Rajasekaran}},
  \bibinfo{journal}{Phys. Rev.} \textbf{\bibinfo{volume}{153}},
  \bibinfo{pages}{1617} (\bibinfo{year}{1967}).

\bibitem[{\citenamefont{Hyodo et~al.}(2008)\citenamefont{Hyodo, Jido, and
  Hosaka}}]{Hyodo:2008xr}
\bibinfo{author}{\bibfnamefont{T.}~\bibnamefont{Hyodo}},
  \bibinfo{author}{\bibfnamefont{D.}~\bibnamefont{Jido}}, \bibnamefont{and}
  \bibinfo{author}{\bibfnamefont{A.}~\bibnamefont{Hosaka}},
  \bibinfo{journal}{Phys. Rev.} \textbf{\bibinfo{volume}{C78}},
  \bibinfo{pages}{025203} (\bibinfo{year}{2008}).

\bibitem[{\citenamefont{Kanada-En'yo and Jido}(2008)}]{KanadaEn'yo:2008wm}
\bibinfo{author}{\bibfnamefont{Y.}~\bibnamefont{Kanada-En'yo}}
  \bibnamefont{and} \bibinfo{author}{\bibfnamefont{D.}~\bibnamefont{Jido}},
  \bibinfo{journal}{Phys. Rev.} \textbf{\bibinfo{volume}{C78}},
  \bibinfo{pages}{025212} (\bibinfo{year}{2008});
%
\bibinfo{author}{\bibfnamefont{D.}~\bibnamefont{Jido}} \bibnamefont{and}
  \bibinfo{author}{\bibfnamefont{Y.}~\bibnamefont{Kanada-En'yo}},
  {\it ibid.}
  \textbf{\bibinfo{volume}{C78}},
  \bibinfo{pages}{035203} (\bibinfo{year}{2008}).

\bibitem[{\citenamefont{Dote et~al.}(2009)\citenamefont{Dote, Hyodo, and
  Weise}}]{Dote:2008hw}
\bibinfo{author}{\bibfnamefont{A.}~\bibnamefont{Dote}},
  \bibinfo{author}{\bibfnamefont{T.}~\bibnamefont{Hyodo}}, \bibnamefont{and}
  \bibinfo{author}{\bibfnamefont{W.}~\bibnamefont{Weise}},
  \bibinfo{journal}{Phys. Rev.} \textbf{\bibinfo{volume}{C79}},
  \bibinfo{pages}{014003} (\bibinfo{year}{2009}).


\bibitem[{\citenamefont{Sekihara et~al.}(2008)\citenamefont{Sekihara, Hyodo,
  and Jido}}]{Sekihara:2008qk}
\bibinfo{author}{\bibfnamefont{T.}~\bibnamefont{Sekihara}},
  \bibinfo{author}{\bibfnamefont{T.}~\bibnamefont{Hyodo}}, \bibnamefont{and}
  \bibinfo{author}{\bibfnamefont{D.}~\bibnamefont{Jido}},
  \bibinfo{journal}{Phys. Lett.} \textbf{\bibinfo{volume}{B669}},
  \bibinfo{pages}{133} (\bibinfo{year}{2008}).

\bibitem[{\citenamefont{Jido et~al.}(2001)\citenamefont{Jido, Oset, and
  Palomar}}]{Jido:2001am}
\bibinfo{author}{\bibfnamefont{D.}~\bibnamefont{Jido}},
  \bibinfo{author}{\bibfnamefont{E.}~\bibnamefont{Oset}}, \bibnamefont{and}
  \bibinfo{author}{\bibfnamefont{J.~E.} \bibnamefont{Palomar}},
  \bibinfo{journal}{Nucl. Phys.} \textbf{\bibinfo{volume}{A694}},
  \bibinfo{pages}{525} (\bibinfo{year}{2001}).

\bibitem[{\citenamefont{Oset and Weise}(1979)}]{Oset:1979bi}
\bibinfo{author}{\bibfnamefont{E.}~\bibnamefont{Oset}} \bibnamefont{and}
  \bibinfo{author}{\bibfnamefont{W.}~\bibnamefont{Weise}},
  \bibinfo{journal}{Nucl. Phys.} \textbf{\bibinfo{volume}{A319}},
  \bibinfo{pages}{477} (\bibinfo{year}{1979}).

\bibitem[{\citenamefont{Braun et~al.}(1977)}]{Braun:1977wd}
\bibinfo{author}{\bibfnamefont{O.}~\bibnamefont{Braun}} \bibnamefont{et~al.},
  \bibinfo{journal}{Nucl. Phys.} \textbf{\bibinfo{volume}{B129}},
  \bibinfo{pages}{1} (\bibinfo{year}{1977}).

\bibitem[{\citenamefont{Jido et~al.}(2003)\citenamefont{Jido, Oller, Oset,
  Ramos, and Meissner}}]{Jido:2003cb}
  \bibinfo{author}{\bibfnamefont{D.}~\bibnamefont{Jido}},
  \bibinfo{author}{\bibfnamefont{J.~A.} \bibnamefont{Oller}},
  \bibinfo{author}{\bibfnamefont{E.}~\bibnamefont{Oset}},
  \bibinfo{author}{\bibfnamefont{A.}~\bibnamefont{Ramos}}, \bibnamefont{and}
  \bibinfo{author}{\bibfnamefont{U.~G.} \bibnamefont{Meissner}},
  \bibinfo{journal}{Nucl. Phys.} \textbf{\bibinfo{volume}{A725}},
  \bibinfo{pages}{181} (\bibinfo{year}{2003}).

\bibitem[{\citenamefont{Jido et~al.}(2009)\citenamefont{Jido, Oset 
  and Sekihara}}]{Jido:2009jf}
  \bibinfo{author}{\bibfnamefont{D.}~\bibnamefont{Jido}}, 
  \bibinfo{author}{\bibfnamefont{E.}~\bibnamefont{Oset}}, \bibnamefont{and} 
  \bibinfo{author}{\bibfnamefont{T.}~\bibnamefont{Sekihara}}, 
  \bibinfo{journal}{arXiv:0904.3410 [nucl-th]}.

\bibitem[{\citenamefont{Tovee et~al.}(1971)}]{Tovee:1971ga}
\bibinfo{author}{\bibfnamefont{D.~N.} \bibnamefont{Tovee}}
  \bibnamefont{et~al.}, \bibinfo{journal}{Nucl. Phys.}
  \textbf{\bibinfo{volume}{B33}}, \bibinfo{pages}{493} (\bibinfo{year}{1971}).

\bibitem[{\citenamefont{Nowak et~al.}(1978)}]{Nowak:1978au}
\bibinfo{author}{\bibfnamefont{R.~J.} \bibnamefont{Nowak}}
  \bibnamefont{et~al.}, \bibinfo{journal}{Nucl. Phys.}
  \textbf{\bibinfo{volume}{B139}}, \bibinfo{pages}{61} (\bibinfo{year}{1978}).

\bibitem[{\citenamefont{Oset et~al.}(2002)\citenamefont{Oset, Ramos, and
  Bennhold}}]{Oset:2001cn}
\bibinfo{author}{\bibfnamefont{E.}~\bibnamefont{Oset}},
  \bibinfo{author}{\bibfnamefont{A.}~\bibnamefont{Ramos}}, \bibnamefont{and}
  \bibinfo{author}{\bibfnamefont{C.}~\bibnamefont{Bennhold}},
  \bibinfo{journal}{Phys. Lett.} \textbf{\bibinfo{volume}{B527}},
  \bibinfo{pages}{99} (\bibinfo{year}{2002}).

\bibitem[{\citenamefont{Hyodo et~al.}(2003)\citenamefont{Hyodo, Nam, Jido, and
  Hosaka}}]{Hyodo:2002pk}
\bibinfo{author}{\bibfnamefont{T.}~\bibnamefont{Hyodo}},
  \bibinfo{author}{\bibfnamefont{S.~I.} \bibnamefont{Nam}},
  \bibinfo{author}{\bibfnamefont{D.}~\bibnamefont{Jido}}, \bibnamefont{and}
  \bibinfo{author}{\bibfnamefont{A.}~\bibnamefont{Hosaka}},
  \bibinfo{journal}{Phys. Rev.} \textbf{\bibinfo{volume}{C68}},
  \bibinfo{pages}{018201} (\bibinfo{year}{2003});
%
  \bibinfo{journal}{Prog. Theor. Phys.} \textbf{\bibinfo{volume}{112}},
  \bibinfo{pages}{73} (\bibinfo{year}{2004}).

\bibitem[{\citenamefont{Waas et~al.}(1996)\citenamefont{Waas, Kaiser, and
  Weise}}]{Waas:1996xh}
\bibinfo{author}{\bibfnamefont{T.}~\bibnamefont{Waas}},
  \bibinfo{author}{\bibfnamefont{N.}~\bibnamefont{Kaiser}}, \bibnamefont{and}
  \bibinfo{author}{\bibfnamefont{W.}~\bibnamefont{Weise}},
  \bibinfo{journal}{Phys. Lett.} \textbf{\bibinfo{volume}{B365}},
  \bibinfo{pages}{12} (\bibinfo{year}{1996});
%
\bibinfo{author}{\bibfnamefont{M.}~\bibnamefont{Lutz}}, 
  {\it ibid.} \textbf{\bibinfo{volume}{B426}}, \bibinfo{pages}{12}
  (\bibinfo{year}{1998});
%
\bibinfo{author}{\bibfnamefont{A.}~\bibnamefont{Ramos}} \bibnamefont{and}
  \bibinfo{author}{\bibfnamefont{E.}~\bibnamefont{Oset}},
  \bibinfo{journal}{Nucl. Phys.} \textbf{\bibinfo{volume}{A671}},
  \bibinfo{pages}{481} (\bibinfo{year}{2000}).

\end{thebibliography}
\end{document}